\documentclass[12pt]{iopart}

\expandafter\let\csname equation*\endcsname\relax
\expandafter\let\csname endequation*\endcsname\relax 

\usepackage{amsfonts,amssymb,amsmath}
\usepackage{bbold}
\usepackage{graphicx}
\usepackage{subfigure}
\usepackage{hyperref}
\usepackage{color}

\DeclareMathOperator{\eul}{\mathrm{e}}
\newcommand{\dd}{\mathrm{d}}
\newcommand{\im}{\mathrm{i}}
\newcommand{\inc}{\mathrm{in}}
\renewcommand{\phi}{\varphi}

\begin{document}
\bibliographystyle{iopart-num}

\title{Incoherent x-ray scattering in single molecule imaging}

\author{J~M~Slowik$^{1,2,3}$, S-K~Son$^{1,3}$, G~Dixit$^{1,3}$, Z~Jurek$^{1,3}$, and R~Santra$^{1,2,3}$}

\address{$^1$ Center for Free-Electron Laser Science, DESY, Hamburg, Germany}
\address{$^2$ Department of Physics, University of Hamburg, Hamburg, Germany}
\address{$^3$ The Hamburg Centre for Ultrafast Imaging, Hamburg, Germany }
\eads{
\mailto{robin.santra@cfel.de}}

\date{\today}



\begin{abstract}
Imaging of the structure of single proteins or other biomolecules with atomic resolution would be enormously beneficial to structural biology.
X-ray free-electron lasers generate highly intense and ultrashort x-ray pulses, providing a route towards imaging of single molecules with atomic resolution.
The information on molecular structure is encoded in the coherent x-ray scattering signal.
In contrast to crystallography there are no Bragg reflections in single molecule imaging, which means the coherent scattering is not enhanced.
Consequently, a background signal from incoherent scattering deteriorates the quality of the coherent scattering signal.
This background signal cannot be easily eliminated because the spectrum of incoherently scattered photons cannot be resolved by usual scattering detectors.
We present an ab initio study of incoherent x-ray scattering from individual carbon atoms, including the electronic radiation damage caused by a highly intense x-ray pulse.
We find that the coherent scattering pattern suffers from a significant incoherent background signal at high resolution.
For high x-ray fluence the background signal becomes even dominating.
Finally, based on the atomic scattering patterns, we present an estimation for the average photon count in single molecule imaging at high resolution. 
By varying the photon energy from 3.5~keV to 15~keV, we find that imaging at higher photon energies may improve the coherent scattering signal quality.
\end{abstract}

\maketitle

\section{Introduction}
Unravelling the structure of bio-macromolecules is essential to comprehend their function at the molecular level and opens novel opportunities to drug design.
X-ray crystallography, however,  which has resolved the majority of currently known protein structures \cite{pdb}, is suffering from the persistent refusal of numerous interesting proteins to form crystals.
There is hope that the emerging x-ray free-electron lasers (XFELs) \cite{emmaFEL, altarelli2006european, ishikawaFEL, allariaFERMI} will release structural biology from this misery.
A recent breakthrough is the determination of a previously unknown biomolecular structure \cite{redecke-nass} from femtosecond nanocrystallography \cite{chapman-nanocryst, spenceREVIEW} with XFEL radiation.
Imaging of non-crystalline single viruses with XFEL radiation has been demonstrated with about 32 nanometre resolution \cite{seibert}.
Ultimately, coherent diffractive imaging with XFELs strives to reveal the structure of individual bio-macromolecules with atomic resolution \cite{gaffney-chapman,quineyREVIEW,bartyREVIEW}.
XFELs feature x-ray pulses with unprecedentedly high fluence and few-femtosecond duration. 
This combination permits one to obtain enough x-ray scattering signal from single molecules, 
while simultaneously outrunning the Coulomb explosion \cite{neutze,chapman2006femtosecond}.
The Coulomb explosion, which is the photoionisation-induced dynamics of atomic positions during and after the x-ray pulse, has been in the focus of interest \cite{neutze,chapman2006femtosecond,hau-riege-damage,jurek-imaging,jurek-dynamics}. 
The degradation of the scattering pattern due to electronic fluctuations caused by ionisation, called electronic radiation damage, has also been analysed \cite{quiney,curwood,lorenz,son}. 

Surprisingly, the unavoidable incoherent x-ray scattering signal in a single molecule imaging experiment has attracted little attention so far.
In contrast to crystallography there are no Bragg reflections in single molecule imaging, which enhance the coherent scattering signal.
The structural information, however, is encoded exclusively in the coherent x-ray scattering pattern.
The usual imaging pixel array detectors are not designed to energetically distinguish the coherent scattering from the incoherent scattering, which is shifted in photon energy.
Incoherent scattering, therefore, degrades the quality of the signal.
(Note that coherent and elastic scattering are used interchangeably throughout this paper, as well as incoherent and inelastic scattering.)
Estimations of the incoherent scattering contribution in hydrodynamic models of carbon clusters have indicated its influence at high resolution \cite{ziaja,jurek}.
Consequently, quantitative understanding of the incoherent scattering signal is indispensable for future experiments, facility design, and development of data processing algorithms. 
 
We present a rigorous ab initio treatment of incoherent scattering under typical single molecule imaging conditions.
We determine the scattering pattern of a carbon atom, including the radiation damage caused by ionisation during the pulse by solving a rate equation model.
We can then dissect the scattering pattern into the coherent scattering signal containing structural information and a background signal from incoherent scattering.
The incoherent scattering can be distinguished into inelastic scattering on bound electrons and inelastic scattering on free (ionised) electrons.
Furthermore, we discuss the dependence of the scattering signal quality on the x-ray fluence and photon energy.
We believe the present work will shine some light on the question which XFEL machine developments are beneficial for imaging \cite{saldin}.

In the next section we introduce the quantum mechanical treatment of inelastic x-ray scattering, as well as the rate equation approach used 
to treat the electronic damage dynamics.
In section~\ref{sec:results} we present and discuss the results of our calculations for the carbon atom.
In section~\ref{sec:molecule} we give an estimate of the average number of scattered photons from a complex biomolecule.
Then we conclude the paper in section~\ref{sec:conclusions}.
Some more detailed discussions are given in two appendices.

\section{Theory of nonresonant x-ray scattering and electronic damage dynamics}
\label{sec:theory}

\paragraph{Nonresonant x-ray scattering:}
We employ a full quantum theory, because inelastic x-ray scattering, also called Compton scattering for short, is a quantum phenomenon \cite{compton,bergstrom,hubbell}.
XFELs provide hard x rays with photon energy up to about 20\,keV, making a nonrelativistic approach sufficient.
We utilise nonrelativistic quantum electrodynamics based on the minimal coupling Hamiltonian \cite{santra} to describe the light-matter interactions.
We use the Coulomb gauge, and employ atomic units unless otherwise stated.
We focus on nonresonant x-ray scattering.
This is justified because light atoms are most abundant in biomolecules, 
and the relevant x-ray energies are far above all near-edge resonant excitations for light atoms.

The nonresonant x-ray scattering follows from the so-called $\mathbf{A}^2$-interaction Hamiltonian
\begin{equation}
 \label{eq:1}
 \hat{H}_{\mathrm{int}} = \frac{\alpha^2}{2} \int \mathrm{d}^3x\, \hat{\mathbf{A}}^2(\mathbf{x})\, \hat{\mathrm{n}}(\mathbf{x}),
\end{equation}
where $\alpha$ is the fine-structure constant, $\hat{\mathbf{A}}$ is the vector potential operator of the radiation field,
and $\hat{\mathrm{n}}(\mathbf{x})$ is the electronic density operator at position $\mathbf{x}$.
We consider the electronic system to be initially in its ground state $|\Psi_0\rangle$, and the incoming x-ray photons to have energy and momentum $(\omega_\inc,\mathbf{k}_\inc)$.
After the scattering event the electronic system will be in the state $|\Psi_F\rangle$, whereas the scattered x-ray photon $(\omega_s,\mathbf{k}_s)$ 
has transferred energy $\omega=\omega_\inc-\omega_s$ and momentum $\mathbf{Q}=\mathbf{k}_\inc-\mathbf{k}_s$ to the electronic system.
Applying Fermi's golden rule yields the double differential scattering cross section (DDSCS) \cite{schuelke}
\begin{align}
 \label{eq:2}
 \frac{\dd^2\sigma}{\dd\Omega_{\mathbf{k}_s}\dd\omega_s} 
&= \left( \frac{\dd\sigma}{\dd\Omega} \right)_{\mathrm{Th}}
   \frac{\omega_s}{\omega_\inc} 
	\sum_F \delta( E_F - E_0 -\omega) 
	\left| \int \dd^3x\, \langle \Psi_F | \hat{\mathrm{n}}(\mathbf{x}) | \Psi_0 \rangle
	\eul^{\im \mathbf{Q}\cdot\mathbf{x}} \right|^2.
\end{align}
Here,  
$\left( \frac{\dd\sigma}{\dd\Omega} \right)_{\mathrm{Th}}
=\alpha^4 \sum_{\lambda_s} |\boldsymbol{\epsilon}^*_{\mathbf{k}_s,\lambda_s} \cdot \boldsymbol{\epsilon}_{\mathbf{k}_\inc,\lambda_\inc } |^2$
denotes the Thomson scattering cross section (TSCS),
where $\boldsymbol{\epsilon}_{\mathbf{k},\lambda}$ is the polarisation vector ($\lambda=1$ or $2$).
Making the customary independent-particle approximation, we can write 
\begin{equation}
 \label{eq:3}
 |\Psi_F\rangle = \hat{c}^\dagger_f \hat{c}_i | \Psi_0 \rangle,
\end{equation}
where we introduce the creation (annihilation) operator $\hat{c}_p^\dagger$ ($\hat{c}_p$) of the spin-orbital $|\phi_p\rangle$ with energy $\varepsilon_p$.
Furthermore, we assume the initial state to be a single Slater determinant $|\Psi_0\rangle = |\Phi_0\rangle$.
Upon expanding the electron density operator $\hat{\mathrm{n}}(\mathbf{x}) = \sum_{p,q} \phi^\dagger_p(\mathbf{x}) \phi_q(\mathbf{x}) \,\hat{c}^\dagger_p \hat{c}_q$, 
we can express the DDSCS in a simplified way 
\begin{align}
 \label{eq:4}
 \frac{\dd^2\sigma}{\dd\Omega_{\mathbf{k}_s}\dd\omega_s} 
&=  \left( \frac{\dd\sigma}{\dd\Omega} \right)_{\mathrm{Th}}
    \frac{\omega_s}{\omega_\inc} 
    \Bigg(
    \delta(\omega) 
    \left| \sum_i\! \int\dd^3x\, \phi_i^\dagger(\mathbf{x})\phi_i(\mathbf{x}) \eul^{\im \mathbf{Q}\cdot \mathbf{x}} \right|^2 \notag\\
&\qquad +
    \sum_f^{\text{unocc.}} \sum_i^{\text{occ.}} \delta( \varepsilon_f - \varepsilon_i - \omega)
    \left| \int \dd^3x\, \phi_f^\dagger(\mathbf{x})  \phi_i(\mathbf{x}) \eul^{\im \mathbf{Q}\cdot\mathbf{x}}  \right|^2 
    \Bigg),
\end{align}
where the index $i$ runs over all occupied spin-orbitals in the initial state $|\Phi_0\rangle$, and the index $f$ over all unoccupied spin-orbitals in $ | \Phi_0 \rangle$.
Keeping the solid angle element $\dd\Omega_{\mathbf{k}_s}$ fixed, the DDSCS determines the spectrum of the scattered radiation.

The photon detectors used in coherent diffractive imaging experiments are not designed to resolve the photon energy \cite{philipp, strueder}.
This means that only the energy integrated DDSCS is experimentally accessible. 
The energy integrated DDSCS then becomes 
\begin{align}
\frac{\dd\sigma}{\dd\Omega_{\mathbf{k}_s}} 
&= \left( \frac{\dd\sigma}{\dd\Omega} \right)_{\mathrm{Th}}
   \Big( | f(\mathbf{Q}) |^2 + S(\mathbf{Q}) \Big), \label{eq:5}
\end{align}
where coherent (elastic) scattering is governed by the form factor
\begin{equation}
 \label{eq:5_f(Q)}
    f(\mathbf{Q}) = \int\dd^3x\, \sum_i^{\text{occ.}} \phi_i^\dagger(\mathbf{x})\phi_i(\mathbf{x}) \eul^{\im \mathbf{Q}\cdot \mathbf{x}} ,
\end{equation}
and the incoherent (inelastic) scattering is characterised by the static structure factor $S(\mathbf{Q})$.
When the energy transfer is small with respect to $\omega_\inc$, the static structure factor can be simplified significantly using the Waller-Hartree approximation \cite{waller-hartree}
(i.e., $\omega_s/\omega_\inc \approx 1 $ and $\mathbf{Q} \approx \mathbf{k}_\inc - k_\inc \mathbf{k}_s/k_s$).
It then reads
\begin{equation}
 \label{eq:5_S(Q)}
 S(\mathbf{Q}) =  Z - \sum_j^{\text{occ.}} \sum_i^{\text{occ.}} \Big| \int\dd^3x\, \phi_j^\dagger(\mathbf{x}) \phi_i(\mathbf{x}) \eul^{\im \mathbf{Q}\cdot \mathbf{x}} \Big|^2,
\end{equation}
where $Z$ denotes the number of electrons in $|\Phi_0\rangle$.
To obtain Eq.~\eqref{eq:5_S(Q)} we have used the completeness relation of the spin-orbitals
$\sum_f^{\text{unocc.}} |\phi_f\rangle \langle \phi_f| = \mathbb{1} - \sum_j^{\text{occ.}} |\phi_j\rangle \langle \phi_j|$.
Remarkably, the final electronic states do not enter in this equation; all summations are carried out over spin-orbitals occupied  in the initial state $|\Phi_0\rangle$.

\paragraph{Electronic damage dynamics:}
We extend the \textsc{xatom} toolkit \cite{son,xatom} to calculate inelastic x-ray scattering cross sections using the full DDSCS expression in
Eq.~\eqref{eq:4} and the Waller-Hartree approach in Eq.~\eqref{eq:5_S(Q)}, see \ref{sec:appendixA} for a comparison of both methods.
For the electronic structure calculations, we employ the Hartree-Fock-Slater model \cite{slater}.
Open-shell systems, like carbon, are treated by averaging over all possible Slater determinants associated with an initial configuration.
Furthermore, \textsc{xatom} calculates photoabsorption cross sections, fluorescence rates, Auger decay rates, 
as well as coherent scattering cross sections \cite{son}.

Our aim is to calculate the x-ray scattering pattern of a carbon atom.
However, the dominating process in an atom irradiated with a highly intense x-ray pulse is the ionising photoabsorption.
Therefore, we have to include the ionisation and relaxation dynamics in our model to obtain realistic results.
For these electronic damage dynamics we employ a rate equation approach \cite{son, santra-rohringer}.
The transitions between the possible electronic configurations $\{I\}$ are represented by a set of coupled rate equations
\begin{equation}
 \label{eq:8}
 \frac{\dd}{\dd t} P_I(t) = \sum_{I'\neq I} \left[ \Gamma_{I'\to I} P_{I'}(t) - \Gamma_{I\to I'} P_I(t) \right],
\end{equation}
where $P_I$ is the population of configuration $I$, and $\Gamma_{I'\to I}$ is the rate for a transition from configuration $I'$ to $I$.
We include photoabsorption processes and accompanying relaxation processes (Auger decay and fluorescence)
in our model.
The population dynamics are dominated by photoabsorption.
In the regime considered here, the photoabsorption cross section is about one order of magnitude larger than the total Compton cross section;
thus we neglect ionisation by Compton scattering.
For example, at $\omega_\inc=10$~keV we obtain for a neutral carbon atom in its ground configuration a total Compton scattering cross section $\sigma_{\mathrm{Compton}}=2.7$~barn (cf. \cite{boeke}) and a total photoabsorption cross section $\sigma_{\mathrm{abs}}=41.6$~barn.
In large molecules or clusters electron impact ionisation can become a major source of ionisation \cite{hau-riege-damage,jurek-dynamics,kai,ziaja-spoel,ziaja-szoeke,ziaja-london,hau-riege-noneq}.
Because we do not treat electron impact ionisation, our present results rather underestimate the background signal from ionised electrons (see below).
We also neglect electron recombination that might attenuate the ionisation \cite{hau-riege-damage,ziaja-wang,ziaja-chapman}.

The free electrons that are created during the pulse by photoionisation or relaxation processes will also contribute to the scattering signal \cite{neutze,jurek,ziaja}. 
In general the free electrons have a highly nonthermal kinetic energy distribution \cite{hau-riege-noneq}.
Assuming plane wave states for the free electrons, coherent scattering from the free electron cloud can occur in the forward direction only.
Keeping in mind that a free electron with 100~eV energy travels about $59~\mathring{\mathrm{A}}/\mathrm{fs}$, it becomes apparent that
the free electron cloud will expand during the pulse to a volume with a radius much larger than $10~\mathring{\mathrm{A}}$.
Consequently, scattering from the free electrons will occur only at very low resolution.
Thus, neglecting the small scattering angle regime, we can neglect coherent scattering from free electrons.
On the other hand, it follows from Eqs.~\eqref{eq:4} and \eqref{eq:5_S(Q)}, that incoherent scattering from free electrons is possible for all 
scattering angles.
Assuming no energy resolution in the detector, the structure factor of the free electrons is given by the number of free electrons.
Finally, the differential scattering signal is
\begin{equation}
 \label{eq:9}
 \frac{\dd \mathcal{I}}{\dd\Omega_{\mathbf{k}_s}} = \int\dd t\, j(t) \sum_I P_I(t) \frac{\dd\sigma}{\dd\Omega_{\mathbf{k}_s}} \Big|_I ,
\end{equation}
where $j(t)$ is the photon flux at time $t$ of the incident x-ray pulse.
The differential scattering cross section of the $I$-th configuration 
contains the coherent and incoherent scattering from bound electrons as given in Eq.~\eqref{eq:5} and the TSCS for Compton scattering from each free electron \cite{chihara1987,chihara2000},
\begin{equation}
 \label{eq:10}
\frac{\dd\sigma}{\dd\Omega_{\mathbf{k}_s}} \Big|_I
=  \left( \frac{\dd\sigma}{\dd\Omega} \right)_{\mathrm{Th}}
   \Big( | f_I(\mathbf{Q}) |^2 + S_I(\mathbf{Q}) + N^{\mathrm{free}}_I \Big).
\end{equation}
Here, $ N^{\mathrm{free}}_I$ denotes the number of free electrons in configuration $I$.
$f_I(\mathbf{Q})$ and $S_I(\mathbf{Q})$ are the atomic form factor and the static structure factor of 
configuration $I$, respectively.  

\section{Scattering from carbon under imaging conditions}
\label{sec:results}

\begin{figure}[h]
\centering
\includegraphics[width=.7\textwidth]{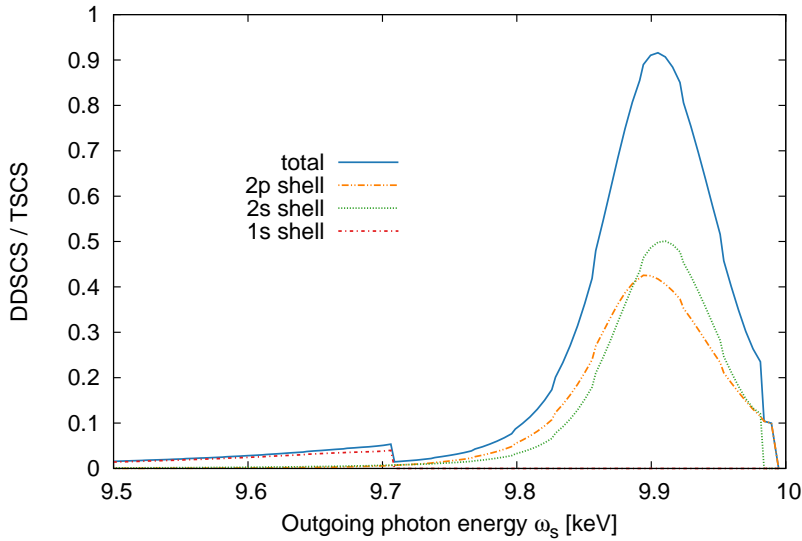}
\caption{The double differential scattering cross section (DDSCS) of a carbon atom in the ground state in units of the Thomson scattering cross section (TSCS) for $\omega_\inc = 10$~keV and $\theta=60^\circ$.
Only the continuous Compton spectrum of ionising bound-to-free transitions is shown; contributions of the atomic orbitals are distinguished.} \label{fig1} 
\end{figure}

We investigate the nonresonant incoherent x-ray scattering from carbon, the most abundant element in biomolecules after hydrogen.
The detector geometry in coherent diffractive imaging experiments at XFELs allows scattering angles up to maximally $\theta \lesssim 70^\circ$ \cite{mancuso,boutet}.
In Fig.~\ref{fig1} we show the spectrum of x rays, scattered incoherently from a neutral carbon atom, for an incoming photon energy of $10$~keV and a scattering angle of $\theta=60^\circ$.
We obtained the spectrum from the DDSC given in Eq.~\eqref{eq:4}.
It shows the continuous Compton spectrum of ionising bound-to-free transitions, at $\theta=60^\circ$ the bound-to-bound transitions are negligible (see \ref{sec:appendixA}).
The spectrum is peaked at a photon energy shift of about 100~eV, and scattering events with an energy shift of more than 200~eV are very unlikely.
As a result it seems to be nearly impossible to filter out the incoherently scattered photons, as most of their spectrum overlaps with the $\sim 0.1\%-1\%$ bandwidth of the x-ray pulse.
This implies that only the energy integrated DDSCS is accessible in experiment.

\begin{figure}[h]
\centering
\includegraphics[width=\textwidth]{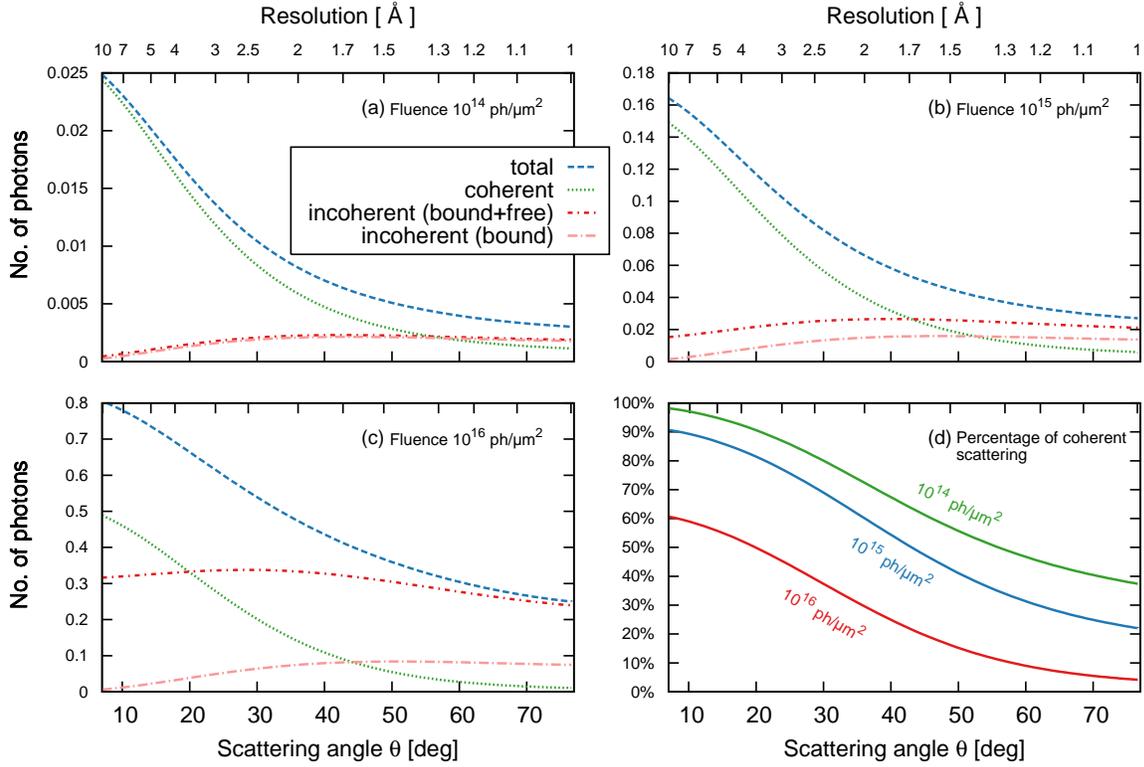}
\caption{The number of photons $\dd\mathcal{I}/\dd\Omega$ scattered from a carbon atom into a solid angle element $\dd\Omega$ for $\omega_\inc=10$~keV.
The x-ray pulse is 10~fs long (flat top) and the fluence is (a) $10^{14}$ photons/$\mu m^2$, (b) $10^{15}$ photons/$\mu m^2$, (c) $10^{16}$ photons/$\mu m^2$. 
The scattering pattern is decomposed into contributions from coherent scattering on bound electrons and the background, i.e., incoherent scattering on bound and free electrons.
Panel (d) shows the percentage of coherently scattered photons in the total scattering.}
\label{fig3} 
\end{figure}

In Fig.~\ref{fig3} we show the scattering pattern of a carbon atom (see Eq.~\eqref{eq:9}) resulting from a 10~fs long, flat top x-ray pulse with $\omega_\inc=10$~keV.
The total scattering pattern is decomposed into a coherent scattering signal and a background signal.
Additionally, the background signal originating from incoherent scattering on bound electrons only is distinguished. 
Fig.~\ref{fig3}(a) shows the scattering pattern for a fluence of $10^{14}$ photons/$\mu m^2$, corresponding to $10^{12}$ photons per pulse focused to $100\times100~\mathrm{nm}^2$, which is currently available at LCLS. 
Observe that the background signal becomes substantial at high resolution.
The contribution of coherently scattered photons drops below 50\% at a scattering angle $\theta\approx55^\circ$, corresponding to $1.36~\mathring{\mathrm{A}}$ resolution.
The background signal is dominantly caused by inelastic scattering from bound electrons, because absorption is not strong enough to strip many electrons off the carbon atom at this fluence.
The mean charge of the carbon atom after the pulse is $+0.75$.
Increasing the fluence to $10^{15}$ photons/$\mu m^2$ one finds an additional background signal from free electrons, see Fig.~\ref{fig3}(b). 
Fig.~\ref{fig3}(c) shows the high fluence case of $10^{16}$ photons/$\mu m^2$, which might be available at future facilities with recently proposed schemes \cite{saldin}.
The scattering pattern dramatically changes in this regime.
The background signal is dominating ($\geq 50\%$) for scattering angles as low as $\theta\approx 20^\circ$ ($3.6~\mathring{\mathrm{A}}$ resolution).
By the end of the pulse most electrons are stripped off the atom at this fluence (the mean charge after the pulse is +5.37).
The dominant contribution to the background signal is caused by incoherent scattering from the free electrons.
At higher scattering angles there is also a strong contribution from inelastic scattering on bound electrons.
Fig.~\ref{fig3}(d) shows the percentage of coherently scattered photons in the total scattering pattern for the three different fluence cases.  

\begin{figure}[ht]
\centering
\subfigure[(a) \& (b) Photon energy {$\omega_\inc=12.4$~keV}.]{\label{fig4a}\includegraphics[width=1.0\textwidth]{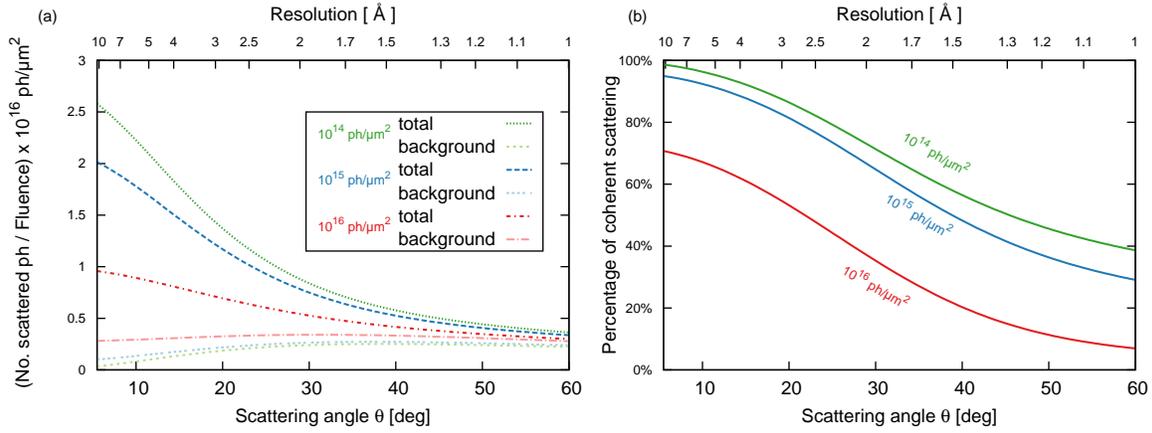} }
\subfigure[(c) \& (d) Photon energy {$\omega_\inc=3.5$~keV}.]{\label{fig4b}\includegraphics[width=1.0\textwidth]{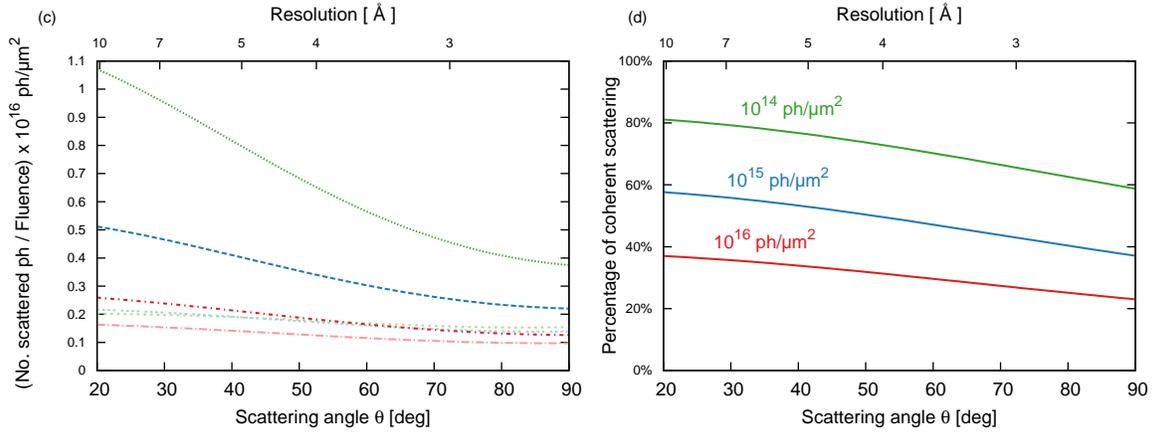} }
\caption{The scattering pattern for $\omega_\inc=12.4$~keV and $\omega_\inc=3.5$~keV.
Panels (a) \& (c) show scattering patterns for fluences of $10^{14}$, $10^{15}$, and $10^{16}$ photons/$\mu m^2$ and 10 fs pulse duration. 
The number of photons is scaled by $10^{16}$/Fluence [in ph/$\mu m^2$] to fit all lines in the same plot. 
The contribution of the background from incoherent scattering is shown.
Panels (b) \& (d) show the percentage of coherently scattered photons.}
\label{fig4} 
\end{figure} 

In Fig.~\ref{fig4} the scattering pattern is presented for photon energies 12.4~keV and 3.5~keV at different fluences.
The pulse duration is 10 fs with a flat top profile.
For 12.4~keV in Fig.~\ref{fig4}(a) we see a substantial degradation at large scattering angles for all fluences. 
Increasing the fluence degrades the quality of the signal also at small scattering angles. 
For a fluence of $10^{14}$~photons/$\mu m^2$ the percentage of the coherent signal drops below 50\% for $\theta \gtrsim 45^\circ$ ($1.3~\mathring{\mathrm{A}}$ resolution), whereas for $10^{16}$~photons/$\mu m^2$ this happens already for $\theta \gtrsim 21^\circ$ ($2.9~\mathring{\mathrm{A}}$ resolution).

It is interesting to note that for 3.5~keV in Fig.~\ref{fig4b}, the dependence on the fluence is more drastic.
The photoabsorption cross section is about 50 times larger than at 12.4~keV and the strong photoionisation fuels the scattering signal from free electrons.
We find that already at a fluence of $10^{14}$~photons/$\mu m^2$ there is a strong background at small scattering angles.
In the forward direction ($\theta=0$) only about $82\%$ of the scattering signal originates from coherent scattering.
However, the coherent scattering signal is prevailing (larger than 60\% up to $\theta \approx 90^\circ$, i.e., $2.7~\mathring{\mathrm{A}}$ resolution).
For $10^{15}$~photons/$\mu m^2$ the coherent scattering signal has a contribution below 60\% and drops below 50\% at $\theta \approx 50^\circ$ ($4.2~\mathring{\mathrm{A}}$ resolution).
At a fluence of $10^{16}$~photons/$\mu m^2$ the background signal is dominating the scattering pattern, and the coherent scattering signal contributes less than 40\% to the scattering pattern throughout the entire range.

\begin{figure}[h]
\centering
\includegraphics[width=\textwidth]{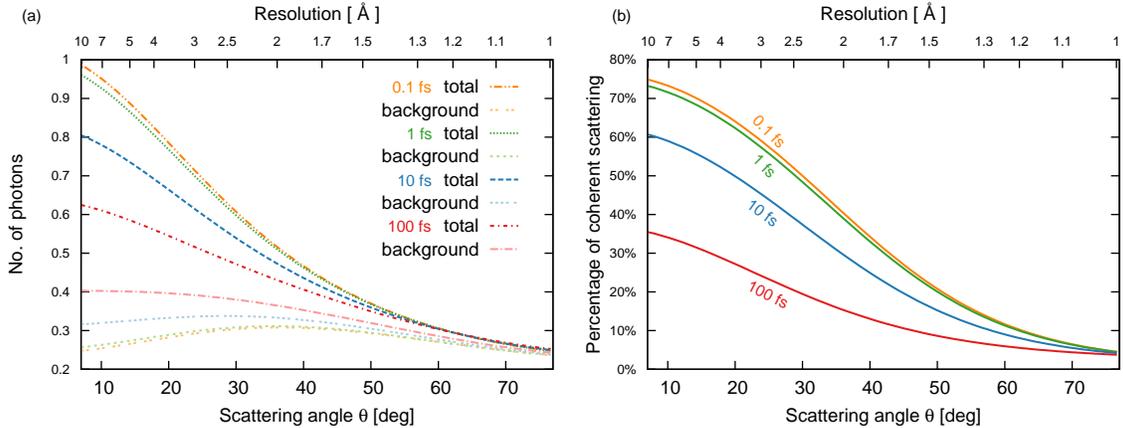}
\caption{Comparison of the scattering pattern for pulses with different pulse duration.
The photon energy is $\omega_\inc=10$~keV, the fluence is $10^{16}$~photons/$\mu m^2$, and the pulse has a flat top profile.} 
\label{fig5} 
\end{figure} 

In Fig.~\ref{fig5} we examine the influence of the pulse duration for $\omega_\inc=10$~keV at $10^{16}$~photons/$\mu m^2$.
The pulse duration is varied from 0.1~fs to 100~fs. 
At large scattering angles or high resolution the scattering pattern suffers from a dominating background irrespective of the pulse duration.
Making the pulse shorter improves the percentage of coherent scattering at small scattering angles or low resolution because of decreased ionisation \cite{young, hoener, son}.
For the present case of a single atom, however, making the pulse shorter than one femtosecond has very little consequence.

\section{Photon counts in single molecule imaging}
\label{sec:molecule}

\begin{figure}[h]
\centering
\subfigure[(a) \& (b) Resolution {$3~\mathring{\mathrm{A}}$}.]{\label{fig6ab}\includegraphics[width=1.0\textwidth]{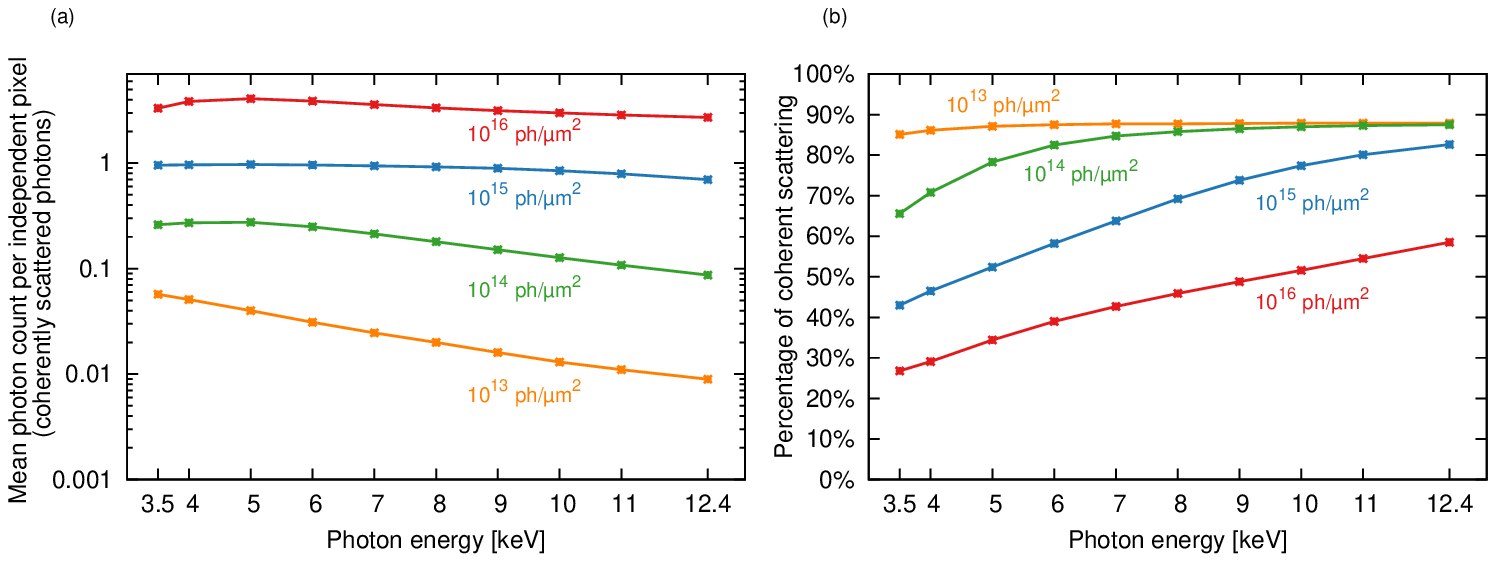} }
\subfigure[(c) \& (d) Resolution {$1.5~\mathring{\mathrm{A}}$}.]{\label{fig6cd}\includegraphics[width=1.0\textwidth]{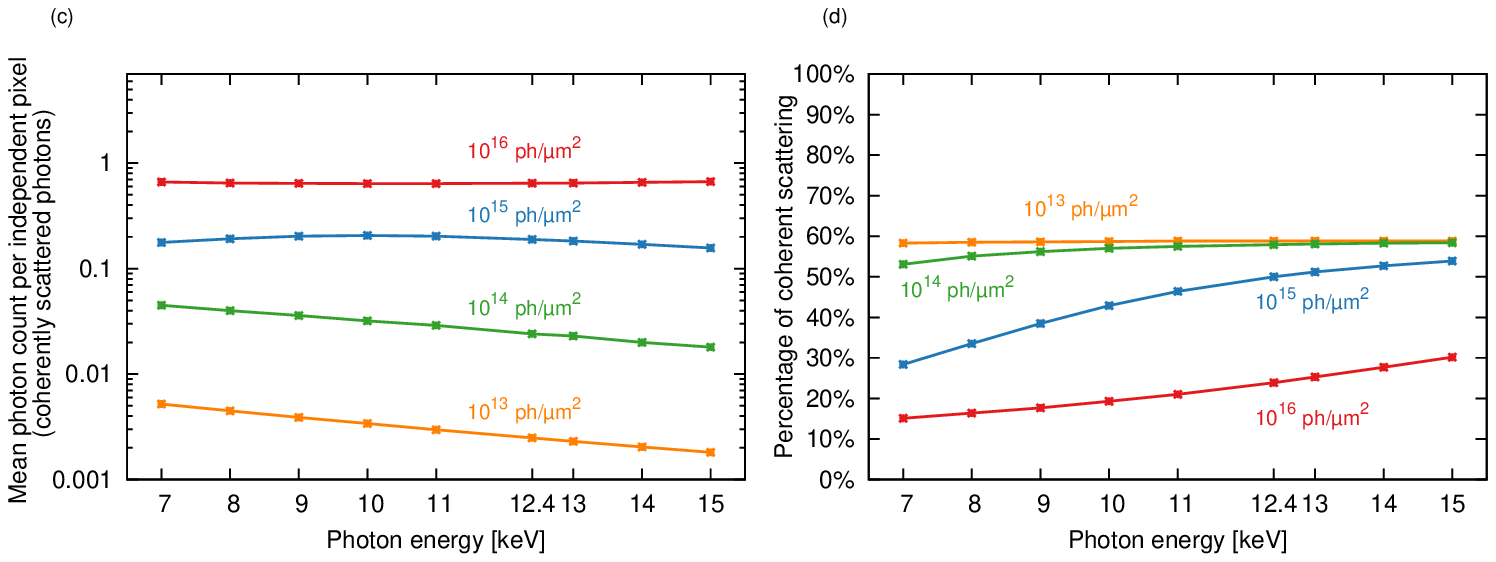} }
\caption{Left panels show the average number of coherently scattered photons 
from a molecule ($\rho_C=1/15~\mathring{\mathrm{A}}^{-3}$, radius $R=100~\mathring{\mathrm{A}}$)
at a fixed resolution, plotted 
versus the incoming photon energy, for four different fluences.
Right panels show the percentage of the coherent scattering signal in the total scattering signal.
The resolution is fixed to $3~\mathring{\mathrm{A}}$ in panels (a) \& (b), and to $1.5~\mathring{\mathrm{A}}$ in panels (c) \& (d).
} \label{fig7} 
\end{figure} 

A typical scheme \cite{huldt, gaffney-chapman} for structure determination of a single molecule is
1) classification of scattering patterns from the same molecular orientations and averaging patterns of the same class,
2) determining the relative orientations of the classes,
3) phase retrieval and reconstruction from the reciprocal space data.
Averaging many images from the same molecular orientation is one way to improve the signal to noise ratio \cite{huldt, bortel2007, bortel2009, bortel2011, tegze2012,yefanov}, but 
there are also methods that merge the classification and orientation into one step \cite{ourmazd,shneerson,loh}.
The challenge for these methods is to deal with the low number of scattered photons.
In Refs.~\cite{huldt,bortel2009} a mean photon count on the order of 0.1~photons/pixel was indicated as necessary for a successful reconstruction with about 2--3~$\mathring{\mathrm{A}}$ resolution.
For small molecules a method that strives to deal with a mean photon count on the order of 0.01~photons/pixel was presented in Ref.~\cite{ourmazd}. 
However, all of these algorithms were developed and tested under simplifying assumptions, in particular neglecting any electronic radiation damage and any background signal. 

We estimate the average number of photons scattered by a single molecule into independent pixels (speckles) at high resolution, taking into account the electronic radiation damage and the background signal.
An independent pixel is the resolution element that corresponds to independent data according to the sampling theorem.
If the molecule fits inside a cube with side length $2R$, an independent pixel collects photons that are detected in a Q-space area of $(\pi/R)^2$ \cite{huldt}.
For x-ray wavelength $\lambda$ the solid angle corresponding to an independent pixel is
\begin{equation}
 \label{eq:4.1}
 \Omega_{\mathcal{P}} = \left( \frac{\lambda}{2 R} \right)^2\;.
\end{equation}
Using very short pulses, we can assume the positions of the atoms to be fixed.
We denote the number of carbon atoms by $N_C$ and the average over independent pixels in the annulus of scattering angle $\theta$ by $\langle\cdot\rangle_{\theta}$.
We assume the scattered intensity into a given independent pixel to be approximately constant over the pixel.
At sufficiently high resolution -- in the regime of Wilson statistics \cite{wilson} -- the average photon number 
\begin{align}
 \label{eq:4.2}
 \left\langle \frac{\dd \mathcal{I}_{\mathrm{mol}}}{\dd\Omega_{\mathcal{P}}} \right\rangle_{\theta}
	&= \Omega_{\mathcal{P}} N_C  \frac{\dd \mathcal{I}}{\dd\Omega_{\theta}} 		
\end{align}
is proportional to the single atom scattering pattern $\dd \mathcal{I}/\dd\Omega_{\theta}$ given in Eq.~\eqref{eq:9}.
(See \ref{sec:appendixB} for details.) 
Note that for a spherical molecule with radius $R$ and carbon atom density $\rho_C$, 
the mean photon count is proportional to $\Omega_{\mathcal{P}} N_C = \frac{\pi}{3} \rho_C \lambda ^2 R \propto N_C^{1/3}$.

In Fig.~\ref{fig7} we assume, as an example, a molecule of $R=10$~nm radius with a carbon atom density of $\rho_C=1/15~\mathring{\mathrm{A}}^{-3}$, 
which is typical for a protein.
We show the average number of photons scattered coherently into independent pixels $\Omega_{\mathcal{P}}$ at fixed resolutions of 
$3~\mathring{\mathrm{A}}$ and $1.5~\mathring{\mathrm{A}}$, as well as the percentage of the coherent scattering.
These numbers are plotted for different photon energies and four x-ray fluences.

Fig.~\ref{fig7}(a) shows the average photon count for $3~\mathring{\mathrm{A}}$ resolution.
It shows that a fluence of $10^{14}$ photons/$\mu m^2$
is sufficient to achieve a mean photon count of 0.1 photons/$\Omega_{\mathcal{P}}$ in the photon energy range of 3--11~keV.
If 0.01 photons/$\Omega_{\mathcal{P}}$ are sufficient for a successful classification of images, the fluence can be lowered
to $10^{13}$ photons/$\mu m^2$.
In practice it may be beneficial to use the lowest necessary fluence.
One may use surplus power in the XFEL beam to increase the focal spot size of the x-ray beam.
A larger focal spot of the x-ray beam improves the hit rate, leading to more efficient use of the XFEL beam and more rapid accumulation of statistically significant data.

Moreover, Fig.~\ref{fig7}(a) displays the effect of radiation damage for imaging.
Without radiation damage the coherent scattering at a given resolution (i.e., at fixed $Q$) does not depend on $\omega_{\inc}$.
Thus one would expect a dependence of the mean photon count per pixel on the photon energy proportional to $\omega_\inc^{-2}$, reflecting the size of $\Omega_{\mathcal{P}}$.
The lowest fluence shows this behaviour.
But the average photon count at the higher fluences is almost constant, with a small peak at 5~keV.
Thus ionisation at high fluence, which is even more pronounced at low photon energy, reduces the coherent scattering.
The influence of radiation damage is also reflected by the fact that an increase of the fluence by an order of magnitude 
does not increase the scattering signal by the same factor.
Furthermore, Fig.~\ref{fig7}(b) shows that at high fluences the percentage of coherent scattering is much lower, due to incoherent scattering on 
ionised electrons.
The incoherent scattering on bound electrons sets an upper limit on the percentage of coherent scattering.
Interestingly, the percentage of coherently scattered photons is much larger at higher photon energies.
Imaging might therefore benefit from using a higher photon energy, because of the improved signal to background ratio.

Similar results are found for imaging with $1.5~\mathring{\mathrm{A}}$ resolution.
Fig.~\ref{fig7}(c) shows that in order to reach 0.1 photons/$\Omega_{\mathcal{P}}$ a fluence of $10^{15}$ photons/$\mu m^2$ is necessary.
The average photon count is not very sensitive to the photon energy, but shows a maximum at 10~keV.
Fig~\ref{fig7}(d) shows that the percentage of coherent scattering at $1.5~\mathring{\mathrm{A}}$ resolution 
is much lower than at $3~\mathring{\mathrm{A}}$ resolution.
In fact, it shows that for $10^{15}$ photons/$\mu m^2$ the background scattering is dominating up to 12.4~keV photon energy,
due to the strong ionisation at this fluence.
Using higher photon energies significantly improves the (fairly low) percentage of coherent scattering.
Improving the signal to background ratio is particularly important because we have completely neglected any noise of the signal.
Consequently, our results suggest that high resolution imaging benefits from the use of photon energies of about $12.4$~keV and higher. 
However, at a fluence of $10^{16}$ photons/$\mu m^2$ there is less than 30\% coherent scattering signal in the total photon count for all photon energies.

\section{Conclusions}
\label{sec:conclusions}

We have determined the background signal from incoherent scattering on bound and free electrons for a single carbon atom under typical imaging conditions. 
The electronic damage dynamics induced by photoabsorption during the pulse have been considered by a rate equation approach.

As a principal result, we have shown that for all considered fluences there is a significant background signal at high resolution.
This background seems to be unavoidable, as the energy shift of inelastically scattered photons is too small to be filtered out.
Even at a moderate fluence of $10^{14}$~photons/$\mu m^2$ at $10$~keV, we find that the background is dominating for scattering angles corresponding to more than $1.3~\mathring{\mathrm{A}}$ resolution.
We have shown that the background signal becomes even stronger at higher fluences due to increased ionisation and subsequent scattering on free electrons.
Scattering on free electrons induces also a strong background at low resolution.
Because electron impact ionisation was neglected in the present study our calculations are somewhat optimistic:
The real number of free electrons may be expected to be larger.
Hence, our findings show that an x-ray fluence of $10^{16}$~photons/$\mu m^2$ is counterproductive for imaging, because the increased scattering on free electrons deteriorates the structural signal.
Thus, our study suggests that strong focusing of high peak power x-ray beams of present and future XFELs seems to be counterproductive for imaging. 
Their strength may rather be an increased hit rate through a larger focal spot size.

Furthermore, we have estimated the average number of photons scattered from a large molecule to high resolution (large scattering angle).
At least for imaging up to $3~\mathring{\mathrm{A}}$ resolution, we found an x-ray beam with a fluence of $10^{14}$ photons/$\mu m^2$ (in the photon energy range 3.5--12~keV)
to scatter sufficiently many photons for a successful classification of molecular orientation.
For imaging up to $3~\mathring{\mathrm{A}}$ resolution a photon energy of 5~keV seems to be ideal.
At higher resolution the background signal becomes significant.
For $1.5~\mathring{\mathrm{A}}$ resolution, we determined a fluence of $10^{15}$ photons/$\mu m^2$ to be necessary for image classification.
At such a high fluence the sample is strongly ionised and x-ray scattering on ionised electrons deteriorate the signal quality.
We have found that the use of photon energies of about 12~keV and higher improves the signal quality at this high resolution.
However, these findings indicate that for imaging at atomic resolution the existing classification algorithms may have to be modified to deal with the background signal.

\appendix

\section{Waller-Hartree vs. energy integrated full DDSCS}
\label{sec:appendixA}

We have implemented in \textsc{xatom} the calculation of the DDSCS given by Eq.~\eqref{eq:4}.
Because we assume no energy resolution, the scattering patterns depend on the static structure factor $S(\mathbf{Q})$,
which is obtained by energy integration of the DDSCS (see Eq.~\eqref{eq:5}).

\begin{figure}[h]
\centering
\includegraphics[width=.7\textwidth]{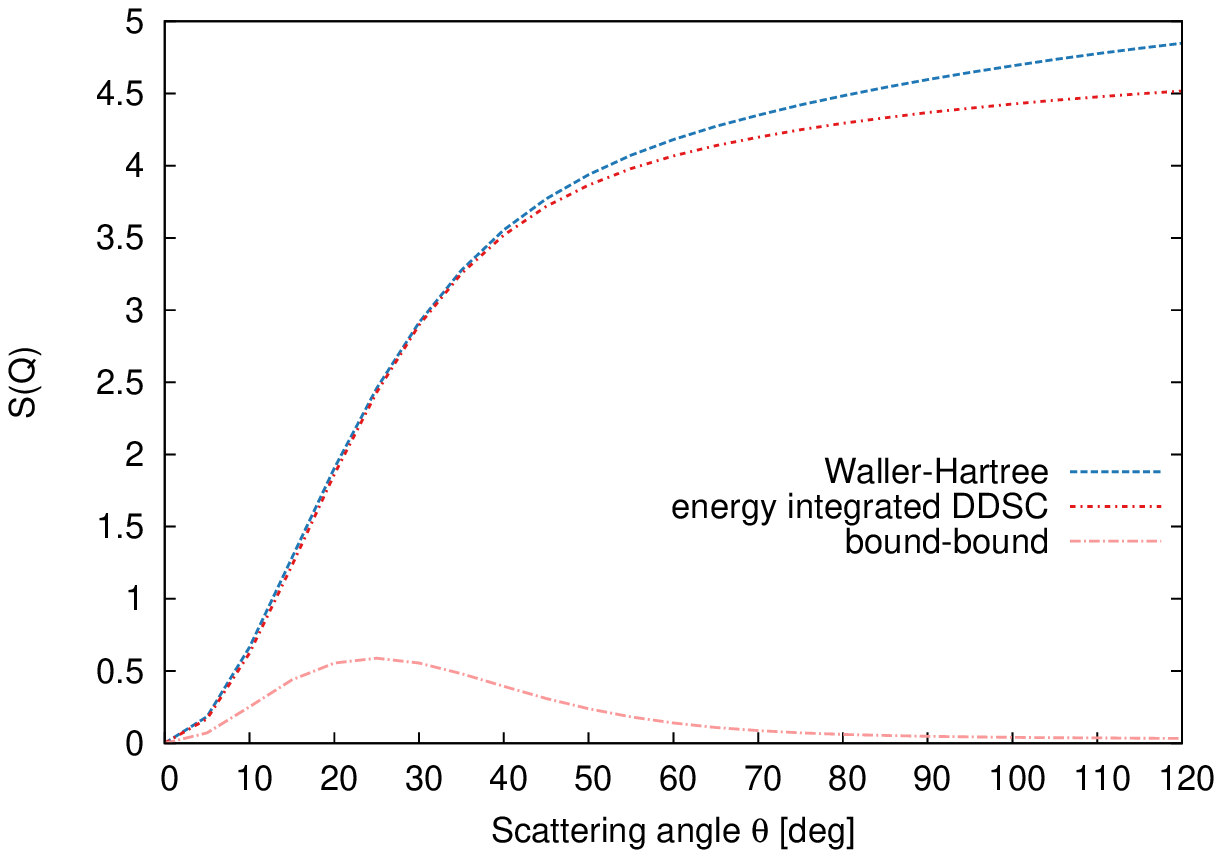}
\caption{The static structure factor for neutral carbon with $\omega_\inc = 10$~keV.
The Waller-Hartree approach is compared with numerical integration of the DDSCS. The contributions of bound-to-bound transitions for the latter method are shown.} \label{fig2} 
\end{figure} 

\noindent $S(\mathbf{Q})$ can be calculated by numerical integration of the incoherent part of the DDSCS
\begin{equation}
\label{eq:A1}
  S(\mathbf{Q})_{\text{full}} = \int \dd\omega \sum_f^{\text{unocc.}} \sum_i^{\text{occ.}} \delta( \varepsilon_f - \varepsilon_i - \omega)
    \left| \int \dd^3x\, \phi_f^\dagger(\mathbf{x})  \phi_i(\mathbf{x}) \eul^{\im \mathbf{Q}\cdot\mathbf{x}}  \right|^2 ,
\end{equation}
or by using the Waller-Hartree approximation (see Eq.~\eqref{eq:5_S(Q)})
\begin{equation}
\label{eq:A2}
 S(\mathbf{Q})_{\text{WH}} =  Z - \sum_j^{\text{occ.}} \sum_i^{\text{occ.}} \Big| \int\dd^3x\, \phi_j^\dagger(\mathbf{x}) \phi_i(\mathbf{x}) \eul^{\im \mathbf{Q}\cdot \mathbf{x}} \Big|^2.
\end{equation}
In Fig.~\ref{fig2} we compare the static structure factor obtained from both methods.
To use Eq.~\eqref{eq:A1} the final states have to be determined, but one can separate the contribution of bound-to-free and bound-to-bound transitions \cite{schnaidt}.
The Waller-Hartree approach is numerically much more favourable.
Because for the relevant scattering angles both approaches are nearly equivalent we have used the Waller-Hartree approximation to perform 
our calculations in the main text.

\section{Molecular scattering pattern}
\label{sec:appendixB}

To a first approximation we can think of complex biomolecules (up to $\sim10^5$ atoms) as a random distribution of $N_{C}$ independent carbon atoms.
For very short pulse duration the atomic positions $\mathbf{R}_i$ are fixed during the pulse.
We assume that scattering at different times adds up incoherently and we average over the populations $P_{\mathbf{I}}(t)$ of the global electronic configurations $\mathbf{I}=(I_1,\dots,I_{N_C})$, where $I_j$ is the electronic configuration of the $j$th atom.
The intensity scattered into a given solid angle $\dd\Omega_{\mathbf{k}_s}$ can be written
\begin{align}
\label{eq:B.1}
 \frac{\dd \mathcal{I}_{\mathrm{mol}}}{\dd\Omega_{\mathbf{k}_s}}
	&= \left( \frac{\dd\sigma}{\dd\Omega} \right)_{\mathrm{Th}}
	\int \dd t\, j(t)
	\sum_{\mathbf{I}} P_{\mathbf{I}}(t) 
	\left[ \; 
	\left| \sum_{j=1}^{N_C} f_{I_j}(\mathbf{Q}) \eul^{\im \mathbf{Q}\cdot\mathbf{R}_j} \right|^2 
	\right. \notag\\
	&\qquad + \left. 
	\sum_{j=1}^{N_C} \Big( S_{I_j}(\mathbf{Q}) + N^{\mathrm{free}}_{I_j} \Big) \; \right]. 
\end{align}
Assuming that the ionisation in one atom is statistically independent of the other atoms, the global population factorises into the individual atomic populations \cite{MADson}
\begin{equation}
\label{eq:B.2}
P_{\mathbf{I}}(t) = \prod_{j=1}^{N_C} P_{I_j}(t). 
\end{equation}
Under these assumptions, as shown in Ref.~\cite{MADson}, the scattering intensity depends only on the single carbon atom population dynamics.
Therefore, we need to consider only electronic configurations $I$ of a single carbon atom, instead of the global configuration~$\mathbf{I}$.

The last term in Eq.~\eqref{eq:B.1} characterises incoherent scattering on bound and ionised electrons.
The incoherent summation over single atoms contains no information on the molecular structure.
The resulting background signal is
\begin{equation}
 \label{eq:B.3}
\frac{\dd \mathcal{I}_{\mathrm{bg}}}{\dd\Omega_{\mathbf{k}_s}}
=
\left( \frac{\dd\sigma}{\dd\Omega} \right)_{\mathrm{Th}}
N_C \int\dd t\, j(t) \sum_{I} P_{I} \Big( S_{I}(\mathbf{Q}) + N^{\mathrm{free}}_{I} \Big).
\end{equation}
On the other hand, the first term in Eq.~\eqref{eq:B.1}  derives from the coherent scattering on bound electrons.
It reduces to
\begin{align}
 \label{eq:B.4}
\frac{\dd \mathcal{I}_{\mathrm{el}}}{\dd\Omega_{\mathbf{k}_s}}
	&= \left( \frac{\dd\sigma}{\dd\Omega} \right)_{\mathrm{Th}} \int \dd t\, j(t)
	\Bigg[ N_C \sum_{I} P_{I} | f_{I}(\mathbf{Q}) |^2 	\notag\\  
	&\quad +
	\Big| \sum_{I} P_{I}(t)  f_{I}(\mathbf{Q}) \Big|^2 \sum_{\substack{i,j\\ i\neq j}} \eul^{\im\mathbf{Q}\cdot(\mathbf{R}_j-\mathbf{R}_i)} 
	\Bigg].
\end{align}
The first term is exactly $N_C$-times the coherent scattering signal of a single atom.
Furthermore, note that $f_I(\mathbf{Q})$ and $S_I(\mathbf{Q})$ are independent of the azimuthal angle and only depend on $\theta$, 
because the atom is always spherically symmetric in our approach.
Thus, the average $\langle \frac{\dd \mathcal{I}}{\dd\Omega_{\mathbf{k}_s}} \rangle_\theta$ only affects the last term in Eq.~\eqref{eq:B.4}.
We consider sufficiently high resolution $r_{\mathrm{res}}=2\pi/Q$.
That is, we assume that $r_{\mathrm{res}}$ is smaller than almost all atom distances $R_{i,j}=|\mathbf{R}_j-\mathbf{R}_i|> r_{\mathrm{res}}$.
For atom distances larger than the resolution, the phase-factor $\mathbf{Q}\cdot(\mathbf{R}_j-\mathbf{R}_i)\gg 2\pi$ oscillates strongly and thus
$\langle \eul^{\im\mathbf{Q}\cdot(\mathbf{R}_j-\mathbf{R}_i)} \rangle_{\theta} = 0 $.
On the other hand, we assume that there are no atom distances $R_{i,j}\ll r_{\mathrm{res}}$, for which the phase factors would add up coherently.
This means, we consider a resolution on the order of the nearest-neighbour distance.
In this case we have random phase factors, and their sum corresponds to the average position of a random walk in the complex plane.
This results in 
\begin{equation}
 \label{eq:B.5}
 \Bigg\langle \sum_{\substack{i,j\\ i\neq j}} \eul^{\im\mathbf{Q}\cdot(\mathbf{R}_j-\mathbf{R}_i)} \Bigg\rangle_{\theta} = 0.
\end{equation}
Hence, at high resolution we can neglect the last term of Eq.~\eqref{eq:B.4}.
In the main text, we considered the case of a molecule with $10$~nm radius and atom density $1/15~\mathring{\mathrm{A}}^{-3}$,
corresponding to $\sim 2.7\cdot10^{5}$ carbon atoms in total.
There are, however, only $\sim 8$ atoms within a radius of $3~\mathring{\mathrm{A}}$ resolution.
The nearest-neighbour distance is $\sim 1.5~\mathring{\mathrm{A}}$.
Assuming that the signal does not vary much over the independent pixel, and combining Eqs.~\eqref{eq:B.3}, \eqref{eq:B.4}, and \eqref{eq:B.5} 
one obtains Eq.~\eqref{eq:4.2}
\begin{align}
 \left\langle \frac{\dd \mathcal{I}_{\mathrm{mol}}}{\dd\Omega_{\mathcal{P}} } \right\rangle_\theta
 &= \Omega_{\mathcal{P}} N_C 
   \left( \frac{\dd\sigma}{\dd\Omega} \right)_{\mathrm{Th}} 
    \int \dd t \, j(t) \sum_I P_I(t) \Big[ |f_I(\mathbf{Q})|^2 +  S_I(\mathbf{Q}) + N_I^{\mathrm{free}} \Big],\\
 &= \Omega_{\mathcal{P}} N_C \frac{\dd \mathcal{I}}{\dd\Omega_{\theta}}.
\end{align}

\ack
Stimulating discussions with Roger Falcone are gratefully acknowledged.
We thank Manish Jung Thapa for valuable input in an early stage of this work.

\vspace{2cm}
\bibliography{bibliography}

\end{document}